\documentclass[sigconf]{acmart}
\AtBeginDocument{%
  }

\copyrightyear{2026}
\acmYear{2026}
\setcopyright{cc}
\setcctype{by}
\acmConference[ICSE '26]{2026 IEEE/ACM 48th International Conference on Software Engineering}{April 12--18, 2026}{Rio de Janeiro, Brazil}
\acmBooktitle{2026 IEEE/ACM 48th International Conference on Software Engineering (ICSE '26), April 12--18, 2026, Rio de Janeiro, Brazil}
\acmPrice{}
\acmDOI{10.1145/3744916.3773150}
\acmISBN{979-8-4007-2025-3/26/04}

\usepackage{colortbl}
\usepackage{pifont}
\usepackage{multirow}
\usepackage{adjustbox} %
\usepackage{colortbl}
\usepackage{xcolor}
\definecolor{LightGray}{gray}{0.95}
\usepackage{pdfpages}   %
\usepackage{makecell}
\usepackage{multirow}
\usepackage{subcaption}

\usepackage{tikz}  %
\usepackage[tikz,usetwoside=true]{mdframed}
\usetikzlibrary{calc,arrows}
\tikzset{
    summary arrow/.style={%
        line width=1pt,
        draw=gray!40,
        rounded corners=1ex,
    },
    summary head/.style={
        fill=white,
        font=\bfseries\sffamily,
        text=gray!80,
        anchor=base west,
    },
}
\mdfdefinestyle{summary}{%
    singleextra={%
        \path let \p1=(P), \p2=(O) in (\x2,\y1) coordinate (Q);
        \path let \p1=(P), \p2=(O) in (\x1,\y2) coordinate (R);
        \path let \p1=(Q), \p2=(O) in (\x1,{(\y1-\y2)/2}) coordinate (M);
        \path [summary arrow] (R) -| (M) |- (P);
        \node [summary head] at ($(Q)+(1ex,-1.5pt)$) {\frametitle};
    },
    firstextra={%
        \path let \p1=(P), \p2=(O) in (\x2,\y1) coordinate (Q);
        \path [summary arrow,latex-] (O) |- (P);
        \node [summary head] at ($(Q)+(1ex,-1.5pt)$) {\frametitle};
    },
    secondextra={%
        \path let \p1=(P), \p2=(O) in (\x2,\y1) coordinate (Q);
        \path let \p1=(P), \p2=(O) in (\x1,\y2) coordinate (R);
        \path [summary arrow] (R) -| (Q);
    },
    middleextra={%
        \path let \p1=(P), \p2=(O) in (\x2,\y1) coordinate (Q);
        \path [summary arrow] (O) -- (Q);
    },
    middlelinewidth=1.5em,middlelinecolor=white,
    hidealllines=true,topline=true,
    innertopmargin=0pt,
    innerbottommargin=1em,
    innerrightmargin=0pt,
    innerleftmargin=1.5ex,
    skipabove=0.5\baselineskip,
    skipbelow=0.3\baselineskip,
}

\ifdefined\DONE
\newcommand{\todo}[1]{}
\newcommand{\dk}[1]{}
\else%
\newcommand{\todo}[1]{\textcolor{red}{TODO:~#1}}

\usepackage[%
    autopunct=true,
]{csquotes} %

\SetBlockThreshold{3}
\usepackage[inline,shortlabels]{enumitem} %
\usepackage{tikz}
\usetikzlibrary{positioning, arrows, calc, backgrounds}
\DeclareQuoteStyle[american]{english} %
        {\itshape\textquotedblleft}
        [\textquotedblleft]
        {\textquotedblright}
        [0.05em]
        {\textquoteleft}
        {\textquoteright}

\newcounter{taskcount}
\begin{document}

\title[Investigating Software Developers' Behavior When Using Sources on Privacy Issues]{{\it``I need to learn better searching tactics for privacy policy laws.''} Investigating Software Developers' Behavior When Using Sources on Privacy Issues}

\author{Stefan Albert Horstmann}
\email{stefan-albert.horstmann@rub.de}
\orcid{0000-0002-4053-0706}
\affiliation{%
  \institution{Ruhr University Bochum}
  \city{Bochum}
  \country{Germany}
}

\author{Sandy Hong}
\email{sandy.hong@rub.de}
\orcid{0009-0002-9206-6893}
\affiliation{%
  \institution{Ruhr University Bochum}
  \city{Bochum}
  \country{Germany}
}

\author{Maziar Niazian}
\email{maziar.niazian@rub.de}
\orcid{0009-0007-3404-3438}
\affiliation{%
  \institution{Ruhr University Bochum}
  \city{Bochum}
  \country{Germany}
}

\author{Cristiana Santos}
\email{c.teixeirasantos@uu.nl}
\orcid{0000-0003-0712-2038}
\affiliation{%
  \institution{Utrecht University}
  \city{Utrecht}
  \country{The Netherlands}
}

\author{Alena Naiakshina}
\email{alena.naiakshina@uni-koeln.de}
\orcid{0009-0008-1843-2027}
\affiliation{%
  \institution{University of Cologne}
  \city{Cologne}
  \country{Germany}
}
\renewcommand{\shortauthors}{Horstmann et al.}
\begin{abstract}
Since the introduction of the European General Data Protection Regulation (GDPR) and the California Consumer Privacy Act (CCPA), software developers increasingly have to make privacy-related decisions during system design and implementation. However, past research showed that they often lack legal expertise and struggle with privacy-compliant development. To shed light on how effective current information sources are in supporting them with privacy-sensitive implementation, we conducted a qualitative study with 30 developers. Participants were presented with a privacy-sensitive scenario and asked to identify privacy issues and suggest measures using their knowledge, online resources, and an AI assistant. We observed developers' decision-making in think-aloud sessions and discussed it in follow-up interviews. We found that participants struggled with all three sources: personal knowledge was insufficient, web content was often too complex, and while AI assistants provided clear and user-tailored responses, they lacked contextual relevance and failed to identify scenario-specific issues. Our study highlights major shortcomings in existing support for privacy-related development tasks. Based on our findings, we discuss the need for more accessible, understandable, and actionable privacy resources for developers.
\end{abstract}



\maketitle

\section{Introduction}
With the introduction of the European General Data Protection Regulation (GDPR)~\cite{GDPR} and the California Consumer Privacy Act (CCPA)~\cite{CCPA}, potential fines~\cite{SaeTheUrb+22}, customer expectations~\cite{horstmann2024those}, and the risk of reputational damage from privacy violations~\cite{CambrideAnalytica} have motivated software companies to ensure privacy-compliant development~\cite{horstmann2024those}. For example, since 2018, over 6 billion Euros have been given out in fines for GDPR violations alone~\cite{enforcementtracker}. As a result, software developers are often tasked with implementing privacy requirements and making privacy-relevant decisions~\cite{horstmann2024those}. 
However, past research found that developers struggle to meet legal requirements, as they often lack privacy expertise~\cite{horstmann2024those, horstmann2025sorry}. While some companies are able to hire legal experts or include privacy champions~\cite{tahaei2021privacy} to support development teams, communication is often limited and challenging~\cite{horstmann2024those, horstmann2025sorry,tahaei2021privacy, tahaei2022embedding}. Consequently, developers frequently have to seek information online independently~\cite{tahaei2022charting, tahaei20understanding}.

With the development of Large-Language Models (LLMs), a novel source to access online legal information has been introduced into the software development process. This shift is reflected in trends, such as Stack Overflow~\cite{stackoverflow} reporting a decline in user activity, which they attributed to the rise of AI systems~\cite{death}. 
The legal information presented in LLMs has the potential to directly
impact present-day developers’ privacy decisions. Incorrect information present in such sources will thus have significant consequences with the usage of these new technologies~\cite{klemmer2024using,leiser24hill,delile23evaluating,kabir2024stack,Tanzil24ChatGPT}. Therefore, it is essential that online sources, including LLMs, provide correct and helpful legal information for current and potential future usage.
To support developers addressing privacy issues during software development, we investigated how they currently seek and use privacy-related information. Further, we examined the challenges they face with existing sources, namely their own knowledge, online sources, and AI assistants, and explored their perspectives on different sources of information, aiming to understand the reasons behind the evolving use of online resources. Through letting the participants use all three sources for a limited time in our study, participants were able to directly compare the different sources and observe the advantages and disadvantages of each.
Lastly, we evaluated these sources to identify how the privacy information available to developers can be improved. %
Thus, we addressed the following research questions:
\begin{enumerate}
    \item[\textbf{RQ1:}] \textbf{What privacy issues do developers identify, and what measures do they propose using their knowledge, online resources, or AI assistance?}
    \item[\textbf{RQ2:}] \textbf{How do developers use and perceive these sources when addressing privacy requirements?}
\end{enumerate}

In this work, we conducted interviews with 30 software development professionals. During think-aloud sessions, participants were presented with a privacy-sensitive scenario and asked to identify privacy issues and possible measures using their knowledge, online sources, and an AI assistant. Afterward, they were asked to reflect on and compare the different sources.

We found that developers prefer using AI assistants over conventional online sources for making privacy-related decisions, as they were perceived as effective and easy to use, despite a general mistrust in the accuracy of these tools. However, our analysis of the AI-generated responses revealed a key limitation: many privacy issues were not reliably identified. The AI assistants often provided vague or generic privacy information and struggled to detect context-specific issues or suggest concrete, scenario-relevant measures.
Based on our findings, we provide recommendations on how online sources can improve their content and make the information more relevant to developers' projects and legal information-seeking.

\section{Related Work}
In this section, we summarize past research on software developers handling privacy requirements, developers using information sources during software development, and the introduction of LLMs into the software development process. 

\subsection{Developers and Legal/Privacy Requirements}
\label{section:DevelopersandLegal}
Responsibility for privacy compliance is often shifted to technical roles, as they are the ones responsible for implementing the requirements~\cite{stover2023website, horstmann2024those}.
However, past research showed ongoing challenges for software developers in understanding and implementing privacy requirements~\cite{senarath2018developers, alhazmi2020developers, alhazmi2021m, horstmann2024those, horstmann2025sorry}. 
For example, research by Senarath et al.~\cite{senarath2018developers} and Alhazmi et al.~\cite{alhazmi2020developers, alhazmi2021m} emphasized that developers lack formal knowledge of privacy principles such as Privacy by Design, Data Minimization, and Privacy Impact Assessments. 
Further, Horstmann et al.~\cite{horstmann2024those} found a communication gap between developers and privacy experts through interviews. In a follow-up lab study~\cite{horstmann2025sorry}, they explored whether access to a privacy expert would support implementation but found that developers were reluctant to make use of it.
Kilhoffer et al.~\cite{kilhoffer2024compliance} also examined how privacy engineers (PE) act as intermediaries between legal and technical teams in large organizations. Their qualitative study found that while PEs help translate legal requirements into actionable technical guidelines, their role is often constrained by organizational priorities, particularly a focus on achieving baseline compliance rather than promoting privacy as a value.
Complementing these findings, Khedkar et al.~\cite{khedkar2024android} conducted an empirical analysis of Android applications, revealing widespread inconsistencies between what developers report in privacy documentation and what is implemented in source code. By analyzing 70 Android apps, their study uncovered both over-reporting and under-reporting of privacy-related data collection in the Google Play Store's~\cite{googleplay} data safety section. 
The study highlights that developers' misunderstandings are compounded by unclear regulatory definitions, insufficient tool support, and a lack of standardized guidance. 

\subsection{Developers and Information Sources}
\label{section:InformationSources}
In the absence of clear internal guidelines or accessible legal expertise, especially in smaller companies with limited resources, developers are often required to seek external information to resolve the privacy issues associated with software development, including forums, blogs, and official documentation~\cite{tahaei2022embedding}. Especially, Stack Overflow~\cite{stackoverflow} serves as a key platform where developers seek assistance on a range of privacy-related topics, including privacy policies, access control, and platform version changes~\cite{tahaei20understanding}. Questions on the platform range from conceptual inquiries in early development phases to direct requests for implementation instructions and help with unexpected errors. While many accepted answers point to official documentation, over half are provided by community members without external references, highlighting a strong reliance on shared peer experience~\cite{tahaei20understanding}. Other forums, such as Reddit~\cite{reddit}, are also used to seek and share information related to privacy regulations. Parsons et al.~\cite{parsons2023understanding} analyzed threads from developer-focused subreddits, identifying common themes and tracking sentiment around GDPR and CCPA. Their findings reveal a strong focus on permission-related questions and regulatory compliance, particularly GDPR. However, the support these traditional sources offer is often inadequate. Previous research identified a lack of high-quality resources, such as educational videos or example codes, that specifically focus on integrating privacy features~\cite{tahaei2022embedding, tahaei2022charting}. Even when resources are available, developers find it hard to navigate and apply, for example, when development platforms mandate the inclusion of a privacy policy but fail to provide guidance on how to create it. Further, in favor of functionality or revenue, software development tools often marginalize privacy considerations~\cite{tahaei2022embedding}. This can manifest itself in ``dark patterns'' in developer interfaces that discourage privacy-friendly options.

\subsection{LLMs in Software Development}
\label{section:LLMsinSoftwareDevelopment}
LLMs have emerged as an innovative and effective source and have quickly incorporated themselves into the daily workflows of software professionals~\cite{kilhoffer2024compliance}.
While prior research mainly focused on code generation and completion~\cite{xu2022systematic, du2024evaluating, yan2022whygen}, some engineers are already using tools such as ChatGPT~\cite{chatgpt-blog-intro} and GitHub Copilot~\cite{copilot}, with promising applications for fine-tuned LLMs that can process privacy documents and automatically extract technical and legal requirements~\cite{kilhoffer2024compliance}, suggesting that the capabilities of these technologies can go beyond simple code generation and completion. Studies showed that developers generally trust code completion tools more than generation tools, as they perceive the suggestions as more precise and think that they are derived from reliable and verified API documentation~\cite{oh2024poisoned}. Still, there is a tendency of developers to ``blindly trust'' AI-generated output without adequate review or testing, which could negatively impact overall software security~\cite{klemmer2024using}. 
In fact, previous research revealed challenges regarding the quality and security of the generated code~\cite{pearce2025asleep, perry2022users, imai2022github}. 
Providing more context or adjusting model parameters can lead to more secure results, while uncritically accepting suggestions can increase security vulnerabilities, underscoring that the tool's effectiveness depends on the users' skill in prompting and critically reviewing the results~\cite{perry2022users}. Additionally, using third-party LLMs introduces other privacy risks. Developers are particularly concerned about possible leakage of sensitive data and intellectual property since the information could be used for training future models, or be exposed in a data breach, leading to developers policing themselves by refraining from inputting actual code~\cite{klemmer2024using}. Further, developers' perceptions are often tied to security and privacy concerns regarding risks associated with closed-source models and external models~\cite{kruse2024can}.
While existing research provides valuable insights into the security implications of AI-assisted code~\cite{asare2024user, sandoval2023lost}, a significant gap exists regarding its impact on privacy compliance, as \enquote{the legal status of code produced by LLMs is an open question}~\cite{sandoval2023lost}.

In comparison to prior research, we extended the current field by examining the state of privacy information through (i) evaluating information sources based on both the results they deliver and their usability, (ii) understanding developer behavior and identifying the challenges associated with each type of source, and (iii) shedding light on the reasons behind the ongoing shift from online sources to AI assistants as a primary source of information.

\section{Research Method}
We conducted and recorded semi-structured interviews with 30 developers. 
Participants were first sent a brief pre-survey to collect demographic information. After completing the survey, they could schedule an appointment using the calendar application Calendly~\cite{calendly}. The interviews were conducted via Zoom~\cite{Zoom}.
As part of the interview, we used a think-aloud approach and asked participants to find privacy issues and measures for a fitness app scenario. 
Finally, we conducted a post-survey on their experience with privacy and the different sources.
Following best practices~\cite{serafini2024engaging}, the parameters of our study are summarized in Table~\ref{table:study_parameters}.
\begin{table}
\caption{Study Parameters.}
\centering
\label{table:study_parameters}
\begin{adjustbox}{max width=\columnwidth}
\begin{tabular}{ll}
\hline
\rowcolor{LightGray} \textbf{Study Type} & Online  \\
\textbf{Study Task}& Think Aloud \& Interview  \\
\rowcolor{LightGray} \textbf{Study Language}& English  \\
\textbf{Length}& mean: 71.79  min (md: 70.12, $\sigma$: 9.82)   \\
\rowcolor{LightGray}\textbf{Recruitment Channel}& Upwork.com \\
\textbf{Recruitment Duration}& 1 month  \\
\rowcolor{LightGray}\textbf{Participants}& Software Professionals (n = 30)  \\
\textbf{Compensation}& \$60 per participant  \\
\hline
\end{tabular}
\end{adjustbox}
\end{table}

\subsection{Scenario}
\label{sec:scenario}
\subsubsection{Description}
Participants were provided a scenario describing the lifecycle of personal data within a hypothetical fitness application. The fitness app scenario was chosen based on related work~\cite{klemmer2023make,horstmann2025sorry}, as it contains a realistic scenario of an application handling sensitive information like health data.
Similar to~\cite{horstmann2025sorry}, we modeled our scenario based on a real fitness application (e.g., Strava~\cite{strava} or Nike Training Club~\cite{nike}). 
The app \textit{collects} user data, including user workout information and, optionally, additional sensitive health data such as blood pressure. 
This data is then \textit{stored} within the system. 
The application \textit{processes} the data to
provide personalized fitness recommendations, suggest sports products from the company’s online store, and display personalized ads to users.
Participants were also informed about the \textit{sharing} and \textit{transfer} of data, including the use of AWS for hosting the application,  and the fact that it would be available worldwide for all common mobile operating systems. 
A detailed scenario description can be found in the supplementary material (see Section~\ref{sec:availability}). We introduced privacy issues into the application, which were based on related work~\cite{horstmann2024those} or real-life GDPR violations~\cite{datareport}.

\subsubsection{Legal Background}
\begin{table}

\caption{Legal Baseline of Scenario.}
\label{tab:legal}
    \begin{tabular}{p{120pt}|p{100pt}}
    Privacy Issue & Suggested Measures\\
\toprule
  \rowcolor{LightGray}   Data Collection&DPIA, Explicit Consent, Encryption \\
  
   Retention Period&  Anonymization after Period\\
  
\rowcolor{LightGray}     Sharing: Location Data&Anonymizing Data\\

    Sharing: Third Party Hosting Service &Data Processing Agreement, Encryption\\

   \rowcolor{LightGray}  Processing: Ads&Explicit Consent, Minimization\\
    
    Processing: Profiling &  Anonymization\\

  \rowcolor{LightGray}   Repurposing: Recommendations& Informed Consent\\
   
    Repurposing: Improvement& Informed Consent\\
   
  \rowcolor{LightGray}   Repurposing: Ads& Informed Consent\\
    
   Transfer to non-EU countries
& Local Storage\\
    \bottomrule

    \end{tabular}
\end{table}

\label{sec:legal_background}
We base the background of the fitness app under the GDPR, %
since it applies extraterritorially to all organizations that offer goods or services to, or monitor the online behavior of, individuals within the EU regardless of where it is established (Art. 3(2)), and considering that the GDPR has influenced the adoption of European-style data protection laws globally (known as the Brussels effect)~\cite{bradford2020brussels, eskhita2024brussels}. 
Considering these data cycle descriptions, jointly with a legal scholar who had 5 years of professional experience as a data protection lawyer, we identified segments of the app description that contained legally relevant information which, if left unaddressed, could result in potential GDPR violations. These passages were then mapped to potential measures to ensure compliance. 
Privacy issues and examples of possible measures are depicted in Table~\ref{tab:legal}. The privacy issues observed in the scenario were: (i) excessive data collection not necessary for the application, (ii) missing a data retention period, (iii) sharing of location data, (iv) sharing data with the hosting provider, (v) processing personal data for ads and (vi) profiling, using data for purposes different form their initial purpose, namely (vii) recommendations, (viii) improvements of the application, (ix) advertising, and lastly, (x) transferring data outside of the EU. Some possible measures to take were to conduct a Data Processing Impact Assessment (DPIA), obtain consent from the data subjects, encrypt data, anonymize personal data, enter into Data Processing Agreements (DPA) with third parties, employ data minimization, and store data inside the EU.

\subsection{Study Procedure}
\subsubsection{Previous Privacy Knowledge} 
At the beginning of the interviews, we probed our participants' knowledge of privacy by asking about its role in their professional lives, the regulations and concepts they were familiar with or had applied, and the sources they had previously consulted.
To establish a baseline across participants, we provided a definition of privacy based on data protection and data privacy, following~\cite{GDPRchapter2,WhatisGDPR}, before the think-aloud sessions (see supplementary material in Section~\ref{sec:availability}).

\subsubsection{Think-Aloud Sessions} 
We asked all participants to address privacy issues in a provided scenario (see Section~\ref{sec:scenario}). To avoid priming effects~\cite{cohn2016priming}, all participants first relied solely on their own knowledge. To address a potential learning effect~\cite{gravetter2011research}, participants were randomly assigned to use either online sources or an AI assistant as their second source. Afterwards, they switched to the remaining source, resulting in a total of three sessions, using every source once. Based on our piloting results (see Section~\ref{sec:piloting}), we opted for 10-minute sessions per source, resulting in a total of 30 minutes:
\begin{enumerate}
    \item \textbf{Own Knowledge:} Participants were first asked to explore privacy issues in the given scenario without any help. This allowed us to examine if and how they would address privacy issues based on their knowledge and experience before consulting external sources or seeking additional help (see Section~\ref{section:DevelopersandLegal});
    \item \textbf{Online Sources:} Past research showed that developers often rely on online search engines, Stack Overflow, and official documentation as sources of information~\cite{acar2016you} (see Section~\ref{section:InformationSources}). Therefore, our participants were asked to use the Web to address the given scenario. They could use a web browser installed on their machine and access any online website they preferred. If they attempted to access an AI assistant, we asked them not to use one during this think-aloud session; 
    \item \textbf{AI-Assistant:} We included AI assistants as an additional information source, as past research suggests they are increasingly replacing traditional online sources for developers~\cite{klemmer2024using, death} (see Section~\ref{section:LLMsinSoftwareDevelopment}). We chose ChatGPT, as it is one of the most widely used AI assistants~\cite{stackoverflowsurvey}. Participants were asked to open ChatGPT in a browser to use as they liked. We logged them into a paid ChatGPT-4 account using our own credentials to ensure participants' anonymity and configured the account not to retain or learn from past conversations. All data was deleted after each participant session.
\end{enumerate} 

We refer to participants who used online sources first as \emph{O}, and those who used an AI assistant first as \emph{A}.
Participants were asked to document all identified issues and proposed privacy measures in an output document.
Additionally, after each session, we asked participants open-ended questions adapted from the System Usability Scale (SUS)~\cite{brooke1996sus}. Since our study did not aim for a quantitative analysis, we chose an open-ended format to gather more detailed and contextual feedback. For example, instead of asking participants to rate their agreement with the statement ``\textit{I found the system unnecessarily complex},'' we asked, ``\textit{What are your thoughts on the complexity of the online sources on privacy?}'' This approach helped us understand how participants perceived the usability of each source, allowing them to explain their reasoning and describe specific challenges. We also asked participants whether they trusted the respective sources and how confident they were in the information they found.
After the three think-aloud sessions, we asked participants to share their opinions on the three sources and compare them. 

\subsubsection{Post-Survey}
After the interviews, participants were sent a short survey concerning their experience with privacy and their perception of the different sources for privacy in the Zoom~\cite{Zoom} chat. 
The full interview guide and survey can be found in the supplementary material (see Section~\ref{sec:availability}).

\subsection{Piloting}
\label{sec:piloting}
We initially designed three different scenarios and planned for participants to solve each one using a different source. The order of scenarios and the assigned information sources were randomized, ensuring that each scenario would be addressed with each source across participants.
We piloted this initial study design with three Computer Science student assistants from our research group. However, we noticed that the time participants spent on the tasks varied strongly based on the source used. Further, participants were influenced by the past scenario, making a comparison between the groups difficult. 
Thus, we created a single scenario and piloted the adapted setup with another researcher from the field of IT security, who had comparable programming experience with the student assistant, to test the setup, framing, questions, and study time.

\subsection{Positionally Statement}
In line with best practices~\cite{Ortloff2023DifferentResearchers, Sannon2022PrivacyResearch}, we include a positionality statement to describe the backgrounds of the researchers involved in this work.
Researchers R1, R2, R4, and R5 have background in usable privacy and security research, whereas R3 has over five years of professional experience as a data protection lawyer.

\subsection{Analysis}
We collected the following data:
(i) audio transcripts of the interviews, 
(ii) search terms used and websites visited by participants when using online sources, 
(iii) participants' chat with the AI assistant containing prompts and replies, 
(iv) privacy issues and the measures that were documented by the participants in an output document, and  
(v) survey results. 
\subsubsection{Qualitative Analysis}
\label{sec:qual}
The analysis of  i) interviews, ii)  used search terms and visited websites, and iii) AI chat was conducted by R1 and R2 using thematic analysis according to Braun and Clarke
~\cite{braun2006using}, using MAXQDA~\cite{maxqda}. 
First, both researchers coded two interviews together, using the results of the legal background as a first basis for coding the privacy issues (see Section~\ref{sec:legal_background}), but deriving the rest of the codes inductively through the data. 
Both researchers then independently coded two different interviews, extending the codebook where needed. 
Afterwards, they met and compared the codebooks, discussed changes, and merged similar codes. 
Each researcher then coded all interviews individually, meeting regularly to compare and discuss the codebook, ordering codes into themes. For example, codes related to online sources being outdated, untrustworthy, and lacking information were combined into the theme that online sources lack in quality.

The same process was repeated to code AI responses, as well as the websites and search terms. 
Lastly, similar to related work~\cite{horstmann2024those, krause2025s, klemmer2024using}, all conflicts were resolved through discussion among researchers, for example, when a coder overlooked a statement in the text or when clarifying distinctions between data minimization and purpose limitation in cases where participants’ descriptions were ambiguous. Thus, we achieved full agreement. We reached saturation after 25 participants, but continued to recruit five additional participants to ensure completeness.
We did not calculate the inter-coder reliability, as it is neither recommended by Braun and Clark nor by other research~\cite{braun-clarke-faqs,braun-clarke-interview, Byrne_2021}.

\subsubsection{Output Documents}
\label{sec:OutputDocuments}
R1, R2, and R3 conducted the analysis of the output documents where participants noted privacy issues and measures they found regarding the app scenario. 
For each output document, we coded  the following: 
\begin{itemize}
    \item  privacy issues noted by participants according to Table~\ref{tab:legal}, 
    \item  measures proposed by  participants to address the observed privacy issues, and 
    \item other suggestions made by the participants not directly connected to any privacy issues of the scenario.
\end{itemize}

The coding was performed on the documents after each step (Own Knowledge, Online Sources, AI), i.e., in the state they were after participants had finished using one of the information sources. By examining the new codes added after each step, we could assess the effectiveness of each source, each new code representing an issue or measure identified through that specific source.

\begin{table}[t]
    \caption{Demographics of the 30 participants.}
    \label{tab:demographics}
    \centering
    \footnotesize
    \renewcommand{\arraystretch}{0.95}
    \setlength{\tabcolsep}{0.66\tabcolsep}
    \setlength{\defaultaddspace}{0.25\defaultaddspace} %
\begin{adjustbox}{max width=\columnwidth}
    \begin{tabular}{lllll}
        \midrule
       \rowcolor{LightGray} \multicolumn{5}{l}{\textbf{Gender}}  \\
       \rowcolor{LightGray} & Man & 24 (80\%) & Woman & 6 (20\%) \\
        \addlinespace
        \multicolumn{5}{l}{\textbf{Age [years]}} \\
        & Min. & 19  & Max. & 67  \\
        & Mean ($\sigma$) & 32.93  ($\pm$9.09) & Median & 31 \\
        \addlinespace
       \rowcolor{LightGray} \multicolumn{5}{l}{\textbf{Experience in current field [years]}}\\ 
       \rowcolor{LightGray} & Min. & 2  & Max. & 24  \\
      \rowcolor{LightGray}  & Mean ($\sigma$:) & 8.52 ($\pm$5.91) & Median & 6 \\
        \addlinespace
        \multicolumn{5}{l}{\textbf{Education}}  \\
         & Bachelor's degree & 12 (40\%) & Master's Degree & 11 (36.67\%)\\
         & High School Equivalent & 2  (6.67\%) & Some College & 2 (6.67\%)\\
         & Vocational Degree & 1  (3.33\%) &Some Grad School&1 (3.33\%)\\
         & Less than high school & 1 (3.33\%) &\\
         \addlinespace
        \rowcolor{LightGray} \multicolumn{5}{l}{\textbf{Current Employment Status}}  \\
        \rowcolor{LightGray} &Full time &14 ( 46.67\%)&Part time  &8 (26.67\%)\\
        \rowcolor{LightGray} &Self-Employed &7 (23.33\%) &Prefer not to say&1 (3.33\%)\\
         \addlinespace
        \multicolumn{5}{l}{\textbf{Largest Company Size (Employees)}}\\ 
        & Min. & 10  & Max. & 300 000  \\
        & Mean ($\sigma$:) & 18760.25 ($\pm$64600)& Median & 135 \\
       \rowcolor{LightGray} \multicolumn{5}{l}{\textbf{Country of Residence}}\\ 
      \rowcolor{LightGray}  &US: 6 &CA: 5&FR: 3&IN: 2\\
      \rowcolor{LightGray}  &UK: 2&Other Europe: 8 &Other Asia: 2&Africa: 2\\
        \bottomrule
    \end{tabular}
\end{adjustbox}
\end{table}

\subsection{Recruitment and Demographics}
We recruited participants through Upwork~\cite{upwork}, as past research~\cite{kaur2022where} recommended it as a suitable recruitment platform for developer studies with practical tasks. 
To address recruitment bias~\cite{delgado2004bias}, we framed the invitation as a research study on information sources used in software development, but did not specifically mention privacy and data protection. 
Participants were required to (1) be 18 years or older, (2) have professionally worked in software development after 2018 to ensure they worked while the GDPR and the CCPA were in force, and (3) describe their last 6 years of software development experience to ensure they were qualified for the study.  
Participants were paid \$60 for the 60-minute study, as aligned with recommendations from~\cite{serafini2024engaging}. 
Overall, Upwork invited 1373 freelancers to apply to the job post. Of those, 378 submitted an application, and Upwork~\cite{upwork} suggested 292 of them. We individually checked all recommended participants' profiles to verify their experience. We contacted 50 freelancers to confirm their continued interest in participating; 30 accepted the invitation. Of the remaining freelancers, two said they were busy with other projects, one was traveling for an extended period, and the remaining 17 did not respond. 
Participants' demographics can be found in aggregated form in Table~\ref{tab:demographics}, with additional information on each participant in Table~\ref{tab:participants}.
Our participants represented 17 different countries: 13 were from Europe, 11 from North America, four from Asia, and two from Africa. Of the 30 participants, 24 identified as men and six as women. Participants had, on average, 8.52 (sd: 5.91) years of experience in their current working domain. 
\begin{table}[t]
\centering
\caption{Overview of the 30 participants.}
\label{tab:participants}
\footnotesize
\setlength{\tabcolsep}{1pt}
\begin{adjustbox}{max width=\textwidth}
\begin{tabular}{llccccc}
\toprule
 & \makecell{\textbf{Current}\\\textbf{Job Title}} 
 & \makecell{\textbf{Ctry.}}&\makecell{\textbf{Company}\\\textbf{Employees}} & \makecell{\textbf{Privacy}\\\textbf{Tasks}}&  \makecell{\textbf{Data Privacy}\\\textbf{Knowledge}}\\ %
\midrule
O1  &Data Scientist&IN &1-99&All  & Very\\ %
\rowcolor{LightGray} O2   & Senior Web Dev & PK&1-99& Half&Moderately\\  %
O3   &Lead Software &US&1000+ &Some &Slightly\\ %
\rowcolor{LightGray} O4   &Full-Stack Dev &LV&1-99 &Half&Slightly\\ %
O5  &Software Dev& TR&100-999&Some &Not at all \\ %
\rowcolor{LightGray} O6   &Software Architect&EE &1000+ &Some & Moderately\\ %
O7  &Full-Stack Dev&US&100-999 &Some &Slightly \\ %
\rowcolor{LightGray} O8   &Software Dev& BA&1-99 &Some &Slightly \\ %
O9  &Software Dev &CA&1-99 &Some &Slightly \\ %
\rowcolor{LightGray} O10  &Senior Dev&VN &1-99 &Some &Not at all\\ %
O11 &Not IT related&CA&100-999 &Some &Moderately \\ %
\rowcolor{LightGray} O12  &Software Dev&US &1-99 &Most &Very\\ %
O13 &Software Dev&CA &1000+ &Some &Moderately \\ %
\rowcolor{LightGray} O14  &Software Dev&GB& 1000+&All &Moderately\\ %
O15  &Software Dev &FR&1-99 &Some &Moderately\\ %
\midrule
\rowcolor{LightGray} A1   &Automation Eng&LT & 1-99&Some&Moderately \\ %
A2   &SecOps Manager&EG &1000+ &Most&Moderately \\ %
\rowcolor{LightGray} A3   &Full-Stack Dev&IN &1-99 &Most &Moderately \\ %
A4   &Software Dev&UA & 100-999&Some &Slightly\\ %
\rowcolor{LightGray} A5   &Software Dev&ET &100-999 &Most &Very \\ %
A6   &IT Cons.&US &1000+ &Half &Moderately \\ %
\rowcolor{LightGray} A7   &Software Eng &CA&100-999 &Some &Not at all\\ %
A8   &Not IT related&US &100-999 &Some &Slightly \\ %
\rowcolor{LightGray} A9   &Senior Dev&CA &100-999 &None &Slightly\\ %
A10  &CTO&GB &1-99 &Most &Very\\ %
\rowcolor{LightGray} A11  &Full-Stack Dev&FR &1-99 &Some &Slightly\\ %
A12  &Web Dev&PT&100-999 &Some &Slightly \\ %
\rowcolor{LightGray} A13  &Full-Stack Dev&FR&100-999 &Some &Slightly\\ %
A14  & IT Cons.&US&100-999 &Some &Not at all \\ %
\rowcolor{LightGray} A15  &Software Devg&ES&1000+ &Some &Slightly \\ %
\bottomrule
\end{tabular}
\end{adjustbox}
\end{table}

\subsection{Ethics}
Our university did not have an Institutional Review Board (IRB) at the time of this study. However, we did comply with the GDPR obligations and followed the principles outlined in the Menlo report~\cite{menlo}. We provided participants with a consent form, which they were asked to download and review before giving their consent to participate. Prior to the interview, we reiterated information about their rights, data collection, and data processing, and gave them the opportunity to ask questions or raise any concerns. Participants could withdraw from the study at any time without consequences. All participants also gave verbal consent at the beginning of the interview recording. The collected data was anonymized after the study, and all recordings were deleted following transcription and verification of data completeness (e.g., ensuring no missing browser history or ChatGPT responses).

\subsection{Threats to Validity}
Our study has several limitations that should be considered when interpreting the results.
First, our participants were freelancers recruited through Upwork~\cite{upwork}, and therefore may not fully represent the global developer population. However, we ensured that each participant had prior experience working in a company setting.
Second, we focus on the GDPR as the primary framework for evaluating participants’ solutions. We chose the GDPR because of its broad applicability to any company handling the data of EU citizens and its global influence through the so-called Brussels Effect, which has shaped the adoption of European-style data protection laws worldwide (see Section~\ref{sec:legal_background}). 
While focusing on other legislation might have revealed additional or different issues, we expect that the GDPR allowed us to identify the core set of privacy issues likely to appear across regulations and countries. While we randomized the order of using the AI assistant and online sources, a learning bias may still have influenced the results.

\section{Results}
\begingroup
\newcommand{\circlepercent}[1]{%
  \begin{tikzpicture}[baseline=-0.5ex]
    \def\radius{0.125cm}
    \pgfmathsetmacro{\angle}{#1 * 3.6}
    \pgfmathsetmacro{\endangle}{90 - \angle}
    \draw[gray] (0,0) circle (\radius);
    \fill[black] (0,0) -- (90:\radius) arc [start angle=90, end angle=\endangle, radius=\radius] -- cycle;
  \end{tikzpicture}%
}

    \renewcommand*{\arraystretch}{1.5}%
    \definecolor{tabred}{RGB}{230,36,0}%
    \definecolor{tabgreen}{RGB}{0,116,21}%
    \definecolor{taborange}{RGB}{255,124,0}%
    \definecolor{tabbrown}{RGB}{171,70,0}%
    \definecolor{tabyellow}{RGB}{255,253,169}%
    \newcommand*{\redtriangle}{\textcolor{tabred}{\ding{115}}}%
    \newcommand*{\greenbullet}{\textcolor{tabgreen}{\ding{108}}}%
    \newcommand*{\orangecirc}{\textcolor{taborange}{\ding{109}}}%
    \newcommand*{\headformat}[1]{{\small#1}}%
    \newcommand*{\vcorr}{%
      \vadjust{\vspace{-\dp\csname @arstrutbox\endcsname}}%
      \global\let\vcorr\relax
    }%
    \newcommand*{\HeadAux}[1]{%
      \multicolumn{1}{@{}r@{}}{%
        \vcorr
        \sbox0{\headformat{\strut #1}}%
        \sbox2{\headformat{\strut #1}}%
        \sbox4{\kern\tabcolsep\redtriangle\kern\tabcolsep}%
        \sbox6{\rotatebox{45}{\rule{0pt}{\dimexpr\ht0+\dp0\relax}}}%
        \sbox0{\raisebox{.5\dimexpr\dp0-\ht0\relax}[0pt][0pt]{\unhcopy0}}%
        \kern.75\wd4 %
        \rlap{%
          \raisebox{.25\wd4}{\rotatebox{45}{\unhcopy0}}%
        }%
        \kern.25\wd4 %
        \ifx\HeadLine Y%
          \dimen0=\dimexpr\wd2+.5\wd4\relax
          \rlap{\rotatebox{45}{\hbox{\vrule width\dimen0 height .4pt}}}%
        \fi
      }%
    }%
    \newcommand*{\head}[1]{\HeadAux{\global\let\HeadLine=Y#1}}%
    \newcommand*{\headNoLine}[1]{\HeadAux{\global\let\HeadLine=N#1}}%
    \noindent
\begin{table}
	\caption{Privacy issues found by participants using different sources}
 \centering
  \setlength{\tabcolsep}{4.2pt}
	\begin{tabular}{lccc}
       &\head{\textbf{Own Knowledge}}&\head{\textbf{Online Sources}}&\head{\textbf{AI Assistant}}\\
		\midrule
    \textbf{Privacy Issues}&&\\
    Data Collection&\circlepercent{53.33}&\circlepercent{6.66}&\circlepercent{56.66}\\
    Retention Period&\circlepercent{0}&\circlepercent{3.33}&\circlepercent{36.66}\\
    Sharing: Location Data&\circlepercent{0}&\circlepercent{3.33}&\circlepercent{10}\\
    Sharing: Third Party Hosting&\circlepercent{36.67}&\circlepercent{3.33}&\circlepercent{23.33}\\
    Processing: Ads&\circlepercent{40}&\circlepercent{0}&\circlepercent{33.33}\\
    Processing: Profiling&\circlepercent{6.67}&\circlepercent{0}&\circlepercent{3.33}\\
    Repurposing: Recommendations&\circlepercent{3.33}&\circlepercent{0}&\circlepercent{0}\\
    Repurposing: Improvement &\circlepercent{10}&\circlepercent{0}&\circlepercent{3.33}\\
    Transfer to non-EU countries &\circlepercent{13.33}&\circlepercent{3.33}&\circlepercent{16.67}\\
        \bottomrule
\end{tabular}
	\label{tab:issues_found}
    \captionsetup[subtable]{labelformat=empty}
    \subcaption{\circlepercent{0} represents the percentage of participants who added the issue to the output document while using the respective source}
\end{table}
\endgroup

In this section, we present the analysis of the interview and think-aloud sessions, participants' output documents, their browsing history (\emph{Browser History}) when using online sources, and their interactions with the AI assistant (\emph{AI Chat}). 
Privacy issues and measures proposed are presented in Section~\ref{sec:sol_doc} and the use of privacy sources is reported in Section~\ref{sec:sources}.
Throughout the section, we differentiate between \textit{reported data}, i.e., thoughts and opinions of the participants, and \textit{observed data}, i.e., sources we saw participants access, and \textit{results} from the analysis of the AI interaction.

\subsection{RQ1: Privacy Issues and Measures}
\label{sec:sol_doc}

We report the effectiveness of the three different sources based on the issues identified and measures proposed by participants in their output documents in Section~\ref{sec:effect}. We then describe privacy issues in Section~\ref{sec:issues} and measures in Section~\ref{sec:solutions} documented in participants’ output documents, as well as those mentioned during interviews and think-aloud sessions. 
While participants have identified issues and measures, they did not always include them in their written responses. This may suggest difficulties in relating the information from the source to their specific use case, or challenges in understanding it. 
Therefore, we distinguish between what was \textit{verbally} expressed and what was \textit{documented}.
While certain issues arose directly from the scenario description (see Table~\ref{tab:legal}), participants also discussed additional common concerns. Further, participants proposed a wide range of measures—some aligning with those outlined in the legal background, while others were independently suggested.  

\subsubsection{Effectiveness of Sources}
\label{sec:effect}
\begingroup
\newcommand{\circlepercent}[1]{%
  \begin{tikzpicture}[baseline=-0.5ex]
    \def\radius{0.125cm}
    \pgfmathsetmacro{\angle}{#1 * 3.6}
    \pgfmathsetmacro{\endangle}{90 - \angle}
    \draw[gray] (0,0) circle (\radius);
    \fill[black] (0,0) -- (90:\radius) arc [start angle=90, end angle=\endangle, radius=\radius] -- cycle;
  \end{tikzpicture}%
}

    \renewcommand*{\arraystretch}{1.5}%
    \definecolor{tabred}{RGB}{230,36,0}%
    \definecolor{tabgreen}{RGB}{0,116,21}%
    \definecolor{taborange}{RGB}{255,124,0}%
    \definecolor{tabbrown}{RGB}{171,70,0}%
    \definecolor{tabyellow}{RGB}{255,253,169}%
    \newcommand*{\redtriangle}{\textcolor{tabred}{\ding{115}}}%
    \newcommand*{\greenbullet}{\textcolor{tabgreen}{\ding{108}}}%
    \newcommand*{\orangecirc}{\textcolor{taborange}{\ding{109}}}%
    \newcommand*{\headformat}[1]{{\small#1}}%
    \newcommand*{\vcorr}{%
      \vadjust{\vspace{-\dp\csname @arstrutbox\endcsname}}%
      \global\let\vcorr\relax
    }%
    \newcommand*{\HeadAux}[1]{%
      \multicolumn{1}{@{}r@{}}{%
        \vcorr
        \sbox0{\headformat{\strut #1}}%
        \sbox2{\headformat{\strut #1}}%
        \sbox4{\kern\tabcolsep\redtriangle\kern\tabcolsep}%
        \sbox6{\rotatebox{45}{\rule{0pt}{\dimexpr\ht0+\dp0\relax}}}%
        \sbox0{\raisebox{.5\dimexpr\dp0-\ht0\relax}[0pt][0pt]{\unhcopy0}}%
        \kern.75\wd4 %
        \rlap{%
          \raisebox{.25\wd4}{\rotatebox{45}{\unhcopy0}}%
        }%
        \kern.25\wd4 %
        \ifx\HeadLine Y%
          \dimen0=\dimexpr\wd2+.5\wd4\relax
          \rlap{\rotatebox{45}{\hbox{\vrule width\dimen0 height .4pt}}}%
        \fi
      }%
    }%
    \newcommand*{\head}[1]{\HeadAux{\global\let\HeadLine=Y#1}}%
    \newcommand*{\headNoLine}[1]{\HeadAux{\global\let\HeadLine=N#1}}%
    \noindent
\begin{table}
	\caption{Measures found by participants using different sources}
 \centering
  \setlength{\tabcolsep}{4.2pt}
    \small
	\begin{tabular}{lccc}
       &\head{\textbf{Own Knowledge}}&\head{\textbf{Online Sources}}&\head{\textbf{AI Assistant}}\\
		\midrule
    \textbf{Measures}&&\\
    Transparency&\circlepercent{40}&\circlepercent{30}&\circlepercent{50}\\
    Consent&\circlepercent{36.67}&\circlepercent{16.67}&\circlepercent{36.67}\\
    Encryption&\circlepercent{20}&\circlepercent{3.33}&\circlepercent{50}\\
    Legal Compliance&\circlepercent{20}&\circlepercent{16.67}&\circlepercent{26.67}\\
    User Deletion&\circlepercent{16.67}&\circlepercent{10}&\circlepercent{46.77}\\
    Data Minimization&\circlepercent{3.33}&\circlepercent{3.33}&\circlepercent{40}\\
    Anonymization&\circlepercent{6.67}&\circlepercent{3.33}&\circlepercent{26.67}\\
    Purpose Limitation&\circlepercent{6.67}&\circlepercent{0}&\circlepercent{13.33}\\
    Two Factor Authentication&\circlepercent{6.67}&\circlepercent{0}&\circlepercent{13.33}\\
    Age verification&\circlepercent{3.33}&\circlepercent{3.33}&\circlepercent{23.33}\\
    Auditing&\circlepercent{3.33}&\circlepercent{0}&\circlepercent{16.67}\\
    
        \bottomrule
\end{tabular}
	\label{tab:found_measures}
    \captionsetup[subtable]{labelformat=empty}
    \subcaption{\circlepercent{0} represents the percentage of participants who added the measure to the output document while using the respective source}
\end{table}
\endgroup

The percentages of participants who identified a specific issue using a particular source are presented in Table~\ref{tab:issues_found}.
Similarly, Table~\ref{tab:found_measures} reports the percentages of participants who suggested measures in their output documents while using each source.
To evaluate the effectiveness of each privacy source, we compared the number of new codes that were generated in our analysis after participants used the respective source. Each code represents either a privacy issue or some form of privacy measure found and added by the participant (see Section~\ref{sec:OutputDocuments}). 
Participants performed similarly when applying their own knowledge as a first source, with a mean of 5.2 (min: 2, med: 5, max: 8, sd: 1.66) for the AI-First and 5.27 (min: 1, med: 5, max: 10, sd: 2.63) for the Online-First participants. Looking at the first source of information used, using the AI performed better compared to the online searchers: The AI-First group had 9.73 (min: 0, med: 6, max: 33, sd: 10.75) new codes on average, whereas the group using online first only gained 2.4 (min: 0, med: 3, max: 8, sd: 2.2) new codes on average. When looking at the third iteration of the output document, using the AI assistant again proves more effective: AI-First achieved an average of 1.2 (min: 0, med: 0, max: 6, sd: 1.93) new codes using online sources, whereas the Online-First group achieved 10 (min: 0, med: 4, max: 74, sd: 18.5) on average with the use of ChatGPT.
Considering the results independent of their order of application, the use of online sources resulted in the fewest new codes, with 1.8 new codes on average (min: 0, med: 1.5, max: 8, sd: 2.12), followed by use of their own knowledge with a mean of 5.23 (min: 1, med: 5, max: 10, sd: 2.16).  Using the AI assistants created the most, with 9.87 new codes on average (min: 0, med: 4.5, max: 74, sd: 14.84). Notably, participants who copied ChatGPT responses into the documents drastically increased the number of coded sections. 

This data suggests that online sources require a significantly longer amount of time to use to find privacy issues and measures compared to the participants' own knowledge and AI ChatGPT. However, the number of issues and measures found does not necessarily mean all requirements or the required measures were found. This is especially a concern with the results obtained using AI assistants, as some participants simply copied ChatGPT's response into the output documents, meaning the issues and measures are highly dependent on the correctness of the AI's output.

\subsubsection{Identified Privacy Issues} 
\label{sec:issues}

Participants were able to detect some of the \textit{privacy issues} presented in the description of the application. %
All participants verbally mentioned the issue of data collection for the application, and 23 added it to the output document. \blockquote[O9]{Yeah, don't collect more than you need.} Twenty-six mentioned personal information, with 23 referencing health data, twenty citing activity data, and 16 discussing location. 
Twenty-seven participants noticed the sharing of data with third-party services as an issue: \blockquote[A5]{So there is a lot of user data being stored on AWS and a lot of it is (...) personal information, age, height, blood pressure, stuff like that.} Thirteen made note of it. 
Further, 15 participants documented targeted ads, which were discussed by 13: \blockquote[A14]{But for including advertising services, showing users personalized ads, that would require its own specific consent agreement.}  
Only nine participants noted data transfers as critical, even though nineteen participants discussed the risks associated with cross-border data flows: \blockquote[A15]{Okay, well, first of all, we are speaking about an old good problem of a US company operating across the globe and hence, EU data has to be stored in EU.} 
Fifteen participants expressed concern about data sharing in general, even when no specific third-party was mentioned. \blockquote[O12]{We don't want to share the data publicly.} This concern was registered by five participants. 
Eight were also concerned about the indefinite storage of data described in the task description: \blockquote[O2]{How long will the data is retained by app owner, and what happened to that owner when you stop using app to delete their account?}  
One participant raised the issue of repurposing of user data for ads and for recommendations in the app in their output document, which twenty-one participants discussed. 
Seven of the participants discussed the repurposing of data for improvement and optimization of the app as something to be considered, but only four further wrote it down in their output document. \blockquote[O7]{Can provide that user data instead of improving, motivating, and optimizing for the future project. Okay, we also should ask for this. Like, can we use the data to improve our services?} 
Four participants discussed profiling and noted it down in the output document: \blockquote[O5]{So we are processing users' data to recommend a personalized user experience. Let's say it's a problem.} 
Eight participants were concerned about the sharing of location data: \blockquote[O9]{Yeah, so this is also another privacy issue is that your location because your location is different people have access to it and you can have some people, for example, like following you or knowing exactly where you where you run or you live.} However, only two noted it down. 
Only one participant explicitly mentioned the sharing of health data, which was discussed by six, as an issue in the output document. 

\subsubsection{Suggested Measures}
\label{sec:solutions}
All participants verbally suggested measures aimed at strengthening user rights, i.e., \blockquote[O9]{essentially ask for consent} or \blockquote[O12]{be transparent, clear.} 
Twenty-one suggested transparency, for example, through a privacy policy in the output document: 
\blockquote[O5]{App needs to have privacy policy.} 
Further, 21 of the participants added to the document the need to obtain user consent to process their data. 
Nineteen also suggested user data access, and to give users the option \blockquote[O5]{to correct or delete their data.} 
In this line, 13 participants wrote down about exercising the users' rights, e.g., the deletion of or access to their data. %
With 26 participants discussing them, 22 noted down additional security measures, often encryption of the data: \blockquote[O3]{Securely receive \& store encrypted PII data when receiving information from the frontend application.} 
Purpose limitation was noted by 16 of the participants in the document as well, with 15 introducing some form of data minimization, which was discussed by 21. 
Eighteen suggested including a retention period, either deleting or anonymizing the data after the period was up, which 12 noted down. 

\subsubsection{Comparison of Participants} As the GDPR is also relevant for developers outside of the EU, as long as they are handling data of EU citizens, we also included participants from outside of the EU. Overall, we had 10 participants currently residing in the EU and the UK, where the legislation is based on the GDPR, and 20 outside. EU and UK participants performed slightly more effectively when exploring privacy issues and measures, suggesting 18.7 (min: 5, med: 14.5, max: 38, sd: 12.5) on average, compared to an average of 15.9 (min: 3, med: 12, max: 78, sd: 15.95) suggestions by the non-EU and UK participants. During the interviews, the results were similar as well in almost all cases. When discussing laws, the GDPR was mentioned proportionally more often by EU and UK participants, with 3.5 mentions on average (min. 0, med: 2, max: 12, sd: 3.54), compared to an average of 2.75 (min: 0, med: 2, max: 10, sd: 2.95) mentions by non-EU participants. Instead, the non-EU participants mentioned a wider variety of different laws, such as HIPAA, CCPA, COPPA, and other regional laws. This issue was also observable in the search terms used when using online sources, as well as the chats with the AI assistant. While the data suggests that participants are more likely to be aware of their local legislation, the GDPR remained the most commonly mentioned legislation in every case. However, the qualitative nature of our work to find significant differences in other cases, such as which issues are found, or measures that were suggested by each group.
Lastly, as participants reported their data protection knowledge, we compared the number of issues found and measures suggested by the participants. Participants reporting below-moderate data protection knowledge found on average 15.19 (min: 3, med: 13, max: 35, sd: 9.02) issues and measures, which was slightly lower than participants reporting average knowledge or higher, who achieved 18.71 (min: 3, med: 11.5, max: 78, sd: 19.57) on average.

\label{sec:isseus_priv}

\label{sec:priv_sol}

\subsection{RQ2: Using Privacy Sources} %
\label{sec:sources}
To evaluate the different sources used by participants, we analyzed their comments during think-aloud sessions as well as their responses to interview questions regarding source usability. Throughout the section, we differentiate between the participants' behavior we were able to observe when interacting with sources, and the thoughts and opinions the participants expressed. We present the results of using their own knowledge in Section~\ref{sec:knowledge}, online sources in Section~\ref{sec:online}, and an AI assistant in Section~\ref{sec:ai}.

\subsubsection{Own Knowledge}
\
\label{sec:knowledge}
\textbf{Behavior:} Participants read the scenario and noted down issues they saw in the task description. 
If they knew a privacy measure, they noted it down as well. Since they did not interact with any other source, the information mainly stems from what the participant mentioned during the think-aloud and interview sessions.

\textbf{Explanations from the Interviews:} 
Ten participants reported frequently relying on their own knowledge of privacy, while nine remarked they would not apply their personal understanding to privacy-related tasks: \blockquote[A7]{Stuff like this should not be covered by developers. It should be covered by actual people who understand this stuff.} 
Five participants explained that they might draw on their knowledge to identify privacy issues in the software they developed: \blockquote[O14]{I try to go through the steps that I would think or usually they occur as a privacy issue.} 
Four participants mentioned they had prior knowledge, e.g., through previous projects, online research, or documentation.

\subsubsection{Online Sources} 
In this section, we examine the participants' usage of online sources, drawing from both their browser history and their self-reported accounts.
\label{sec:online}

\textbf{Behavior:}
We examined the online searches and sources that participants accessed. While 28 participants used Google~\cite{google} as their search engine, two participants opted to use DuckDuckGo~\cite{duckduckgo}.
Concerning search terms, 23 of the participants looked for specific issues related to the context of the fitness tracking app, e.g., \blockquote[O13 Browser History]{Potential privacy issues for fitness tracking app in development}, or more specific questions, for example for data storage \blockquote[A4 Browser History]{how data storing is organized in fitness app} or general privacy measures they could take: \blockquote[O8 Browser History]{privacy measurements when building a workout app with stores}. 
Fifteen of the participants looked for answers to specific questions they had, e.g., on health data: \blockquote[A1 Browser History]{can you store patient data in the cloud}, or about advertisements: \blockquote[O4 Browser History]{how to make sure personalized ads are compliant}.
Again, 15 participants specifically looked for information related to laws, for example, the GDPR: \blockquote[A5 Browser History]{gdpr regulation requirements for healthcare}, or similar guidelines: \blockquote[A7 Browser History]{ftc data privacy}. 
Four specifically looked for other fitness applications they knew, e.g., \blockquote[O9 Browser History]{strava privacy issues}. 
Three looked for information from specific online forum sources, for example \blockquote[A15 Browser History]{privacy and data security health reddit} and \blockquote[O5 Browser History]{processing location data of users stackoverflow}. 
Six participants looked for privacy requirements from a specific country, mainly the United States: \blockquote[A4 Browser History]{requirements in us for storing privacy data}, and two participants also looked for information regarding requirements in Germany. 

Additionally, we collected the websites that participants accessed during the study for privacy information. Twenty-six participants accessed blogposts from privacy-focused organizations, e.g.,~\cite{privacyrights,iclg}, companies~\cite{kaspersky,scrut,usercentrics}, or news organizations~\cite{forbes,times}. 
Seventeen visited company sites and checked their privacy policy for information, e.g., Uber~\cite{uber} or Whoop~\cite{whoop}. 
Only seven visited government websites, such as the Federal Trade Commission~\cite{ftc}, the European Council~\cite{consilium}, or the Information Commissioner's Office~\cite{ico}. 
Seven also got information from community sources, such as Stack Overflow~\cite{stackoverflow} or Wikipedia~\cite{wikipedia}. 
Six accessed academic papers during the online research, with three participants accessing a paper on privacy in fitness applications, which suggested blockchain technology as a remedy~\cite{alhajri2022privacy}. Other papers accessed by a participant were related to cloud storage in fitness applications~\cite{zheng2022design}, privacy management for fitness applications~\cite{abdelhamid2021fitness}, and on developers' compliance processes for children's applications~\cite {alomar2022developers}. 
Lastly, one developer accessed information on the Spring framework for security information~\cite{spring}.

\textbf{Explanations from the Interviews:} Fifteen of the participants claimed they would use online sources for privacy decisions, and eight explicitly stated they would not use them. Six stated they would use such sources to find measures for identified issues, and three mentioned they would use them to find information on privacy issues. 
Eight said they would simply use Google %
to find information, with 11 explicitly stating they would look for projects similar to the scenario they are working on, and checking their privacy policy or statements and commitments the companies have made regarding privacy: \blockquote[A12]{I think I need to find another fitness app and to check their privacy, to find out maybe I missed something and they found it and already described in their privacy. It should be some famous app.}
Eight would use online sources to research applicable laws: \blockquote[O4]{I guess I'm just going to Google about GDPR summary and maybe I can find something about that.} 
Two mentioned they would use privacy policy generators, i.e., online tools that automatically generate privacy policies for software or websites (e.g., from Shopify~\cite{shopifypolicygenerator}).

\subsubsection{AI Assistants}
\label{sec:ai}

\textbf{Behavior:}
We examined how participants interacted with ChatGPT. Twenty-one of them provided the assistant with information about the scenario, either by copying sections from or summarizing the task description: \blockquote[A9 AI Chat]{I am building an application which tracks users fitness and will be saving this information with a cloud service provider (ie, AWS). The data gathered will be fitness related (ie, activity) and will also monitor and use the users information (ie, age, height, etc) to determine any recommendations. I need to ensure privacy policies are adhered to. Can you give me a high-level understanding of how I can achieve this?} 
Eleven asked for general privacy information, e.g., \blockquote[O9 AI Chat]{What are typical worries or concerns around privacy in fitness tracking applications?} 
Fourteen asked for information on specific questions, e.g., 
on potential measures: \blockquote[O6 AI Chat]{will oauth2 technology help with data privacy?}; 
on issues they had notice themselves: \blockquote[A14 AI Chat]{If the app includes advertisement services, showing users personalized ads, what are potential locations this data could become exploited? And what are the guidelines to prevent that?}; or 
on legal questions: \blockquote[O6 AI Chat]{how to comply with modern data privacy laws? Describe 5 main points}. 
Three participants employed prompt engineering to improve the results from the AI: \blockquote[O11 AI Chat]{You are expired software developer with knowledge about data privacy on US market. I’ll ask questions next message} 
One participant copied a paper from Alhajri et al.~\cite{alhajri2022privacy}, found online, discussing blockchains as a measure for privacy risks in fitness apps, and asked the assistant to summarize it. Consequently, this was the only time when the assistant started discussing measures using blockchain: \blockquote[O13 AI Chat]{**Blockchain Application**: Using **smart contracts on a blockchain** could automate these consent grants and revocations, ensuring that changes are transparent, auditable, and enforceable.}, showing that the users' prompts can highly influence the AI assistants' output. 
One participant asked if they could trust the AI assistant with privacy questions, to which it replied: \blockquote[O16 AI Chat]{Yes, you can trust these suggestions regarding data privacy. They are based on widely accepted industry standards, best practices, and regulatory frameworks for data security and privacy.}
Our analysis of the AI assistant’s responses revealed that GDPR was the most commonly referenced law by the AI assistant, cited 128 times. This was followed by 91 mentions of CCPA, 53 of HIPAA, and 28 of COPPA.

Regarding \textit{privacy issues} identified by the AI assistant, all responses addressed data collection, with twenty-seven specifically mentioning health data. Twenty-four responses referred to sensitive and personal data in general, and 19 also noted location data. 
Children’s data appeared in 15 of the responses, while 11 mentioned personally identifiable information (PII) and 12 mentioned activity data. Biometric data was cited in five cases, and four responses included references to financial data.
Fifteen of the replies specifically mentioned targeted ads: \blockquote[O5 AI Chat]{Provide options to opt-in or opt-out of data sharing and personalized ads.} 
Further, ten replies discussed profiling. %
With regard to data sharing, twenty-five responses raised concerns about the sharing of data with third parties, such as the hosting service. 
Similarly, data transfer across country borders was a concern in twenty-two replies. %
Six responses mentioned the sharing of location data, and three mentioned health data. 
Lastly, the majority of responses mentioned the general sharing of data as an issue.
Concerning data usage, replies addressed the repurposing of data as an issue. Seventeen addressed advertising and 12 recommendations. %
Further, the missing retention period was criticized in 24 replies. %
Six replies also mentioned repurposing the data for improvements and optimization as a potential issue.
The AI assistant also mentioned issues not directly covered by the app scenario. Twenty-eight discussed data leaks and unauthorized access. 

With respect to the \textit{measures} described by the assistant, all suggested measures related to user rights, and stressed the importance of consent and transparency. 
Next, all replies suggested giving their users control over their data. 
As the AI assistants had mentioned the issue of underage users before, 15 suggested additional safeguards, i.e., obtaining consent from an adult or age verification before the app could be used. 
Ten AI assistants suggested educating their users on the privacy settings. 
In 28 replies, the AI assistant strongly recommended complying with relevant legislation. 
Further, 28 stressed the need to employ data minimization. 
Introducing a retention period was suggested by 25. %
The assistant advised 16 times that companies should evaluate and vet third parties before engaging with them.
Nine suggested Standard Contractual Clauses for data transfer outside of the EU. 
Eleven specifically recommended using privacy by design when building the software. 
Sixteen recommended ensuring purpose limitation, and 24 suggested access control mechanisms.
Further, 25 replies recommended doing regular security or privacy audits. 
Seven responses emphasized the importance of the company staying up to date with legal developments and educating its team on privacy matters. Eleven also recommended contracting a lawyer or hiring a Data Protection Officer to ensure compliance.

\textbf{Explanations from the Interviews:}
Participants reported asking the AI specific questions or giving it a summary of the task.
Only three participants talked about employing some form of prompt engineering, for example, O9: \blockquote[O9 AI Chat]{Also, give a more technical response that would help a developer make appropriate changes}.
Participants went through the list of issues or measures the AI provided:\blockquote[A13]{I'll just see like for each blue point, provide me a measures as it is in the prompt and if I decide to like rework some stuff, I'll just like rework it, feed it to him again.} 
Fourteen of the participants said they had not used AI for privacy before, mainly because those systems are new: \blockquote[O7]{Because it was more than two years ago and at that time ChatGPT wasn't available for the users probably.}
However, nineteen said they have or would use these tools in real-life: \blockquote[O7]{Not yet, but I would use them if I have this task as of today.}

\subsubsection{Comparison of Sources}
\label{sec:comp}

\begin{figure}
    \centering
    \includegraphics[width=\columnwidth]{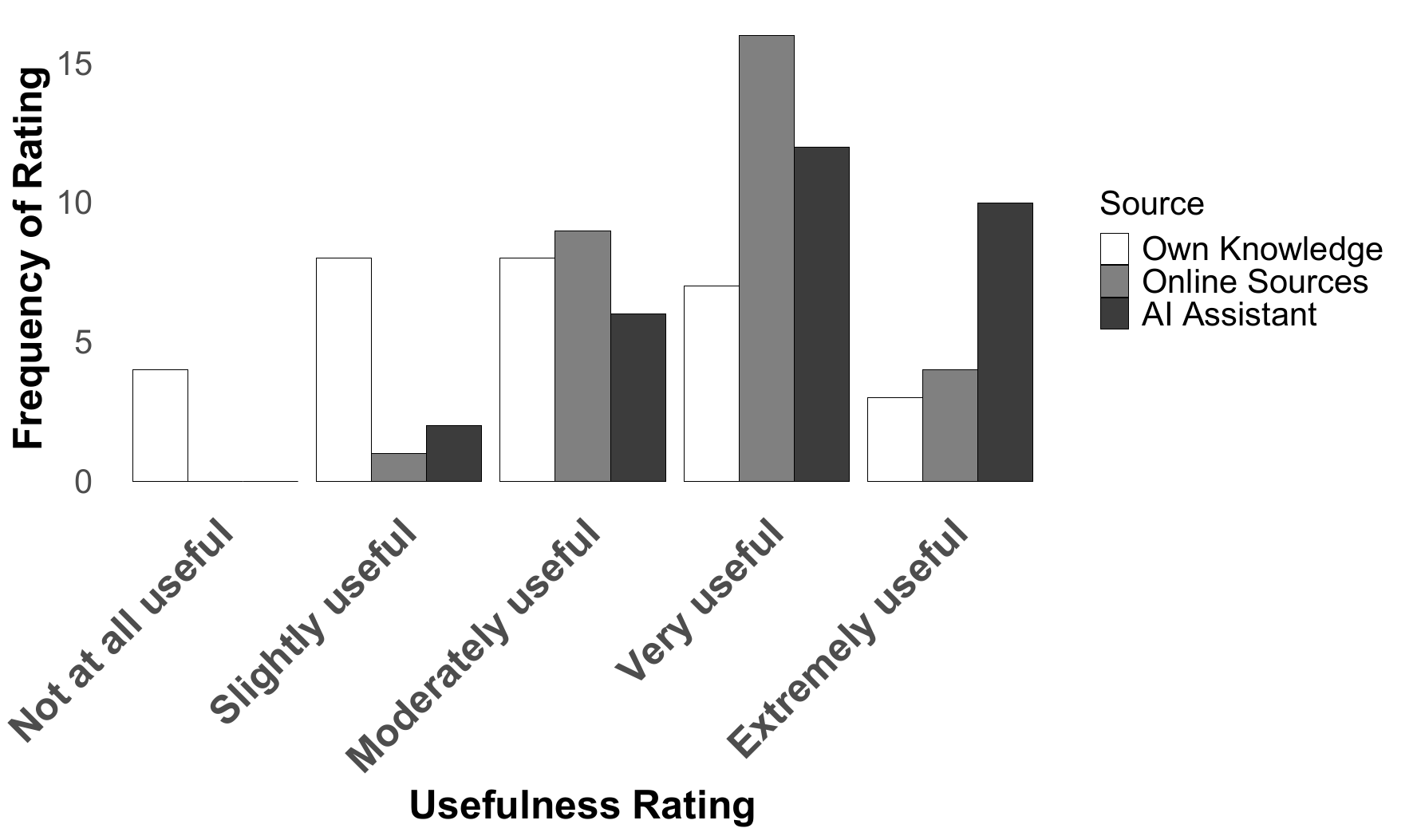}
    \caption{Participants' Rating of the Three Sources}
    \label{fig:rating}
    \Description{Bar graph showing the frequency of the rating of the three different sources, ranging from `Not at all useful' to `Extremely useful'. On average, AI assistants were rated highly most often, followed by online sources. Own knowledge was rated the lowest. }
\end{figure}

A detailed overview of strengths and challenges reported by participants for each source can be found in the supplementary material (see Section~\ref{sec:availability}). 
In the \textit{post-survey}, participants rated the usefulness of the different sources. 
The full distribution of ratings can be found in Figure~\ref{fig:rating}. 
AI assistants were considered the most useful information source, with 10 participants rating them as 'Extremely Useful' and twelve as 'Very Useful.' 
Online sources followed, receiving 4 'Extremely Useful' and 16 'Very Useful' ratings. 
Participants’ own knowledge was rated the least useful, with only three marking it as 'Extremely Useful' and seven as 'Very Useful.'

In the \textit{interviews}, 14 of the participants explicitly said they preferred using AI assistants for privacy research: \blockquote[O10]{Well, I mean, it's not 100 percent easy, but compared in contrast to Google and searching myself. It's much easier than Google.} 
Eight found AI assistants to perform significantly better than online sources: \blockquote[A1]{I think I got more done with ChatGPT, and I was generally more confident with the solutions from ChatGPT in this case. Online, I was really lost, and I did find something, and I don't know if it's even relevant.} 
Six felt AI assistants and online sources performed similarly: \blockquote[A9]{This is kind of going down the same road that ChatGPT went. Where's recommending superiority and privacy and whatever. This one, see like this one came exactly out of ChatGPT. Well, ChatGPT stole this probably from this page.} One participant said that AI assistants have started to replace expert advice: \blockquote[A1]{Yeah, I think before, we relied on experts more.}
Seven participants preferred online sources, mainly as they saw it as the most trusted source: \blockquote[O12]{It would take you more work, I suppose, to search it through general Googling and that sort of thing or whatever, but you could find probably a lot more information, a lot more trusted information and just doing a little extra work.} Four participants believed using their own knowledge for such tasks would yield good results, but noted requiring a much experience: \blockquote[O7]{If the task working with this privacy policies concepts is something you do a lot frequently, you can rely on your experience.} %

Concerning the use of the sources, 17 participants imagined using them together: \blockquote[A5]{I suggest using the three of them because it's ideal to know how the AI will respond to you and what knowledge you had before. And like what you searched online or like Googled or something, then you can collect all this information and bring the best output of what you desired to do.} 

\section{Discussion}
While our study setup had our participants use each source for a relatively short time, our research does offer fewer insights into single sources; however, it does allow insights to directly compare the participants' perception and use of each source. During the interview, participants often mentioned using the three sources in combination with each other. Thus, our setup is similar to how developers would employ these sources at work.

Overall, developers preferred using AI assistants over relying on online sources or their own knowledge. Although participants acknowledged that some understanding of privacy was required to use both AI assistants and online sources effectively, they generally did not expect to complete privacy-related tasks based solely on their own knowledge. This indicates a strong reliance on external support.
When using online sources, developers encountered several challenges in locating and applying relevant information. They frequently found blog posts that addressed isolated issues and were uncertain whether the content was applicable to their specific context. In addition, they considered many sources difficult to interpret and lacking the depth needed to comprehensively cover privacy requirements.
By contrast, the adoption of AI assistants appeared to be primarily driven by efficiency. Developers valued their speed, structured responses, and ease of use. Despite persistent concerns regarding the trustworthiness and accuracy of AI-generated content, these perceived advantages made AI assistants the preferred choice for many participants.

\textbf{Privacy issues.} While participants reliably identified certain privacy issues, particularly those related to data collection and targeted advertising, many concerns remained unaddressed, even with the aid of online resources and an AI assistant. 
Issues involving data repurposing and sharing of location data were especially difficult for participants to detect. 
In particular, repurposing entails knowledge of the GDPR purpose limitation principle (Art. 5(1)(b)) and the need for another legal basis in case further processing is incompatible with the initial purpose for which data was collected~\cite{WP29-2013-PurposeLimitation}. When developers are unable to identify such complex issues, this can lead to unlawful processing and potential breaches.
These issues could not easily be resolved by the sources tested in this study, as online sources were often limited to blog posts and online forums, discussing a few concerns, but were not able to accurately cover privacy issues related to a concrete case, as confirmed in related work~\cite{kyi2025turning}.
Similarly, AI assistants tended to offer generic best practices and rarely identified scenario-specific issues unless explicitly prompted by participants, lacking contextual and legal relevance.
 
\textbf{Measures.} 
While participants were able to recognize some privacy issues and propose standard mitigation measures based on their own knowledge, developers cannot be expected to act as legal experts. 
Participants’ proposed measures were often limited to general best practices and rarely mapped directly to specific legal issues, as outlined in Section~\ref{sec:legal_background}. 
Moreover, the lack of sufficient documentation, often resulting from reliance on informal or unverifiable sources, can hinder the effective implementation of privacy by design principles, as required under Art. 25 of the GDPR. It also weakens an organization’s ability to demonstrate compliance, a core component of the accountability principle under Art 5(2).

Our work shows similarities with related work on information sources for software development and security sources; however, we observe some factors unique to privacy. Different from related work on Stack Overflow~\cite{stackoverflow}, where the main issues concerned the quality of the provided information~\cite{acar2016you,fischer2017stack,fischer2019stack}, our participants had difficulties understanding the information they found and applying it to their own projects. Regarding privacy advice on Stack Overflow, past research showed that developers often rely on online sources for a variety of privacy questions, with many of the answers on Stack Overflow lacking links to official sources, making it challenging to evaluate the quality of replies without any additional knowledge~\cite{tahaei20understanding}. Our results confirm findings from research on privacy during the software development process, showing privacy as a challenging topic for software developers~\cite{horstmann2024those, horstmann2025sorry}. Similar to related work~\cite{liang2024large,klemmer2024using}, our research showed developers favoring the use of AI assistants for their convenience and effectiveness, even though participants reported some degree of mistrust regarding the correctness of the assistants.

Our results suggest that the current state of all three privacy sources does not sufficiently equip developers to independently identify or address significant privacy concerns, making professional legal support necessary, although such support remains infrequently available~\cite{horstmann2024those}.
In the absence of accessible and tailored legal guidance, and given that direct engagement with legal professionals is both rare and sometimes perceived as adversarial~\cite{horstmann2024those, horstmann2025sorry}, developers can be expected to continue relying on alternative sources of support, particularly online resources and AI-based assistants, to meet privacy requirements.
Therefore, ongoing research and policy development are needed to help address the limitations of currently available sources and better support developers in handling privacy-related challenges as follows:
\begin{itemize}
    \item \textbf{Improve Accessibility and Clarity of Official Documentation.} To better support developers during software development, privacy-related information must be more accessible and easier to understand. Legislators could contribute by publishing clear and developer-friendly guidance alongside issued legislation. In its current form, official legal documentations are often avoided, as participants consider it too difficult to interpret and apply.
    \item \textbf{Provide Practical, Jurisdiction-Specific Privacy Guidance.} Developers need reliable and actionable guidance to identify potential privacy issues in their software when navigating privacy requirements independently. Ideally, such guidance should include concrete, implementable scenarios tailored to the jurisdictions relevant to their user base. However, resources of this kind remain relatively limited~\cite{CNIL-2020-DeveloperGuide}. To ensure practical usability and legal accuracy, these should be developed through close collaboration between industry experts and regulatory authorities.
    \item \textbf{Clarify the Role and Limitations of AI Assistants.} AI assistants should disclaim their limitations clearly, particularly when used for sensitive and/or legally relevant privacy tasks (including special categories of data, profiling). Developers must be critically aware of these limitations, verify the accuracy of AI-generated content, and consult additional and reliable sources to ensure comprehensive coverage of privacy requirements.
    \item \textbf{Curated Knowledge Bases.} AI assistants used in legal and privacy contexts must be curated and fine-tuned on legally vetted sources, as in~\cite{leschanowsky2025expert} to ensure trustworthiness and compliance of the information they provide and avoid surface recommendations merely based on popularity, documentation visibility, or commercial marketing materials. 
   \item \textbf{Implication for Industry.} Our results show that, with currently available resources, privacy compliance is challenging for developers, if not supported. Companies should encourage support for open communication with privacy experts, and encourage exchange with more experienced employees to support their development teams. 
    \item \textbf{Future Research.} Future research, both from a legal and computer science perspective, should explore how legal information can be presented in an understandable and actionable way. Further, the integration of AI assistants into the software development process as a primary source of information needs to be carefully studied, as our results already show some of their limitations. 
\end{itemize}

\section{Conclusion}
As privacy grows in importance for legal compliance, developers are increasingly tasked to integrate privacy into software design. However, due to the high cost and limited availability of legal advice, developers frequently address privacy issues independently.
In our study with 30 software professionals, we observed a shift from traditional online searches toward the use of AI assistants as a primary source of information, even for complex and sensitive privacy-related questions. Despite this trend, participants consistently struggled to detect privacy issues and propose appropriate mitigation measures. In many cases, they failed to identify critical issues, and the measures they proposed were often limited to general best practices rather than being tailored to the specific context.
To address these shortcomings, we recommend that privacy guidance be presented in clear, accessible language and closely aligned with the practical realities of software development. While AI assistants can offer helpful insights and identify some issues, they should not be relied upon as the sole source of support, as many relevant concerns remain unaddressed in their responses.

\section{Availability}
\label{sec:availability}
To support the reproducibility and transparency of our paper, we provide the following supplementary material: (i) pre- and (ii) post-surveys, (iii) the interview guide, (iv) the task description given to the participants, and (v) the codebooks resulting from our analysis. To preserve our participants' privacy, we do not share the transcriptions of the recordings. The supplementary material can be found here: https://doi.org/10.6084/m9.figshare.30529859

\begin{acks}
    Funded by the Deutsche Forschungsgemeinschaft (DFG, German Research Foundation) under Germany's Excellence Strategy - EXC 2092 CASA - 390781972 and the RENFORCE research program of the School of Law of Utrecht University.
\end{acks}

\bibliographystyle{ACM-Reference-Format}
\bibliography{./bib}{}

\end{document}